\documentclass{aa}
\usepackage{graphicx}

\begin{document}

\def\spose#1{\hbox to 0pt{#1\hss}}
\def\lta{\mathrel{\spose{\lower 3pt\hbox{$\mathchar"218$}}
     \raise 2.0pt\hbox{$\mathchar"13C$}}}
\def\gta{\mathrel{\spose{\lower 3pt\hbox{$\mathchar"218$}}
     \raise 2.0pt\hbox{$\mathchar"13E$}}}
\def\Msun{{\rm M}_\odot}
\def\msun{{\rm M}_\odot}
\def\Rsun{{\rm R}_\odot}
\def\Lsun{{\rm L}_\odot}
\def\half{{1\over2}}
\def\RL{R_{\rm L}}
\def\zs{\zeta_{s}}
\def\zR{\zeta_{\rm R}}
\def\dJJ{{\dot J\over J}}
\def\dMM{{\dot M_2\over M_2}}
\def\tKH{t_{\rm KH}}
\def\eck#1{\left\lbrack #1 \right\rbrack}
\def\rund#1{\left( #1 \right)}
\def\wave#1{\left\lbrace #1 \right\rbrace}
\def\dd{{\rm d}}

\title{The High-Frequency QPOs in GRS~1915+105}

\author{T. Belloni\inst{1}
	\and
	M. M\'endez\inst{2,3}
	\and
	C. S\'anchez-Fern\'andez\inst{4}
}

\offprints{T. Belloni}

\institute{Osservatorio Astronomico di Brera,
	Via E. Bianchi 46, I-23807 Merate (LC), Italy\\
	\email{belloni@merate.mi.astro.it}
   \and
        Astronomical Institute ``A. Pannekoek'' and Center for High-Energy
	Astrophysics, University of Amsterdam, Kruislaan 403, 
	1098 SJ Amsterdam, the Netherlands\\ 
        \email{mariano@astro.uva.nl}
   \and 
	Facultad de Ciencias Astron\'omicas y Geof\'\i sicas,
	Universidad Nacional de La Plata, Paseo del Bosque S/N,
	1900 La Plata, Argentina\\ 
        \email{mmendez@fcaglp.unlp.edu.ar}
   \and
        Laboratorio de Astrof\'\i sica Espacial 
         y F\'\i sica Fundamental (LAEFF-INTA),
	P.O. Box 50727, E-28080, Madrid, Spain\\
	\email{celia@laeff.esa.es}
}

\date{Accepted for publication in Astronomy \& Astrophysics, 2 April 2001}

\abstract{
We analyzed two 1996 observations of the Galactic Microquasar GRS~1915+105 
taken with the Proportional Counter Array on board the Rossi X-ray
Timing Explorer. We focus on the properties of the 
high-frequency QPO as a function of the `dip' oscillations, a factor of
1000 slower. In one observation, 1996 May 5th, we find that the energy
spectrum of the QPO at $\sim$65 Hz changes drastically 
between the high-intervals
and the dips, hardening during the dips to the point that the QPO peak
is no longer detected at low energies. In the second observation, 1996 May
14th, although it has similar overall characteristics, 
the 65 Hz QPO is not seen, but another broader QPO peak appears
at 27 Hz, only during the dips. This peak is too weak to obtain reliable
information about its spectrum.
Our results indicate that the presence/absence of high-frequency features
in this enigmatic source 
are intimately linked to the slower oscillations and
variations that happen on longer time scales.
\keywords{accretion: accretion disks -- black hole physics -- stars:
	black hole candidates: oscillations -- X-rays: stars}
} 

\maketitle

\section{Introduction}

The X-ray source GRS~1915+105 was discovered in 1992 with WATCH (Castro-Tirado
et al. 1992) as a bright transient (Castro-Tirado et al. 1994). Since then
it has remained bright and active. It was the first galactic object to show
superluminal expansion in radio observations (Mirabel \& Rodr\'\i guez 1994).
It is considered a black-hole candidate because of its high X-ray luminosity
and of its similarity with GRO~J1655-40, for which a dynamical estimate 
of the mass of the compact object has been obtained (Bailyn et al. 1995).

A very large and valuable database of observations has been obtained
with the Rossi X-ray Timing Explorer (RXTE). It is clear from RXTE data that
the source shows a level of variability not seen in any other system
(Greiner et al. 1996; Belloni et al. 1997a,1997b; Chen et al. 1997;
Muno et al. 1999).
Belloni et al. (1997a, 1997b) interpreted this variability as the result
of an instability that causes the innermost region of the accretion disk
to disappear and be refilled again on a viscous time scale. These events
are clearly associated with radio activity and jet ejection (Fender et al.
1997; Eikenberry et al. 1998; Mirabel et al. 1998;
Fender \& Pooley 1998; Klein-Wolt et al., in preparation).
Belloni et al. (2000) reduced the modes of variability into a number of
separate classes, and concluded that all variations could be reduced to
the alternation of three basic states, called A, B and C. In their
interpretation, the major events are caused by B--C transitions, with 
C being the state where the instability is at work and B the state
where the full inner accretion disk is observable.
The new state, state A, seems to be a 
separate third state where the accretion disk is fully
observable, but softer than in state B.

The complexity of this source extends to the fast aperiodic variability.
In the RXTE data, besides strong band-limited noise components, three 
different types of Quasi Periodic Oscillations (QPO) have been
observed. Low-frequency
QPOs at frequencies $\leq$0.1 Hz are observed (Morgan et al.
1997, hereafter MRG97), obviously associated with the
long time scale variations caused by state changes (see Belloni et al. 2000).
Intermediate-frequency QPOs, traditionally called 1-10 Hz QPOs 
are also observed.
They are seen only during state C intervals, are strongly correlated with 
the source flux and energy spectrum, and show a complex phase-lag structure
(MRG97; Markwardt et al. 1999; Reig et al. 2000). These 1-10 Hz
QPOs show strong similarities with other QPOs observed in BHCs (see
van der Klis 1995).
In addition, a QPO at a rather constant frequency of 65-67 Hz has been
observed in some observations (MRG97). Because its centroid frequency was
seen to vary only by a couple of Hertz, this QPO is thought to be 
associated with
a basic frequency of the system, such as the Keplerian frequency at the
innermost stable orbit (MRG97), the Lense-Thirring precession at the same
orbit (Cui et al. 1998), 
or a diskoseismic {\it g-mode} of the disk (Nowak et al. 1997).
Other characteristics are its relatively high $Q$ value (defined as
$\nu / \Delta\nu$) of
$\sim$15 (MRG97), its hard spectrum (MRG97) and its hard phase lags (Cui 1999).
This feature has a fractional rms of only a few \% and seems to be 
detected only during states A and B, being therefore mutually exclusive
with the 1-10 Hz QPO (see MRG97). Indeed, not all detections reported
by MRG97 are very significant, and a full search for the presence of this
QPO has not been yet published.

In this paper, we analyze two RXTE observations of GRS~1915+105 from 1996
May, where the 67~Hz QPO has been detected. In these observations
strong oscillations with a time-scale of $\sim$10 s are observed (see
MRG97), corresponding to oscillations between state A and state B (see
Belloni et al. 2000). We explore the relation between the high-frequency 
QPO and these low-frequency state changes.

\section{Data analysis}

\subsection{Selection of observations}

\begin{figure}
\centering
\includegraphics[width=0.5\textwidth]{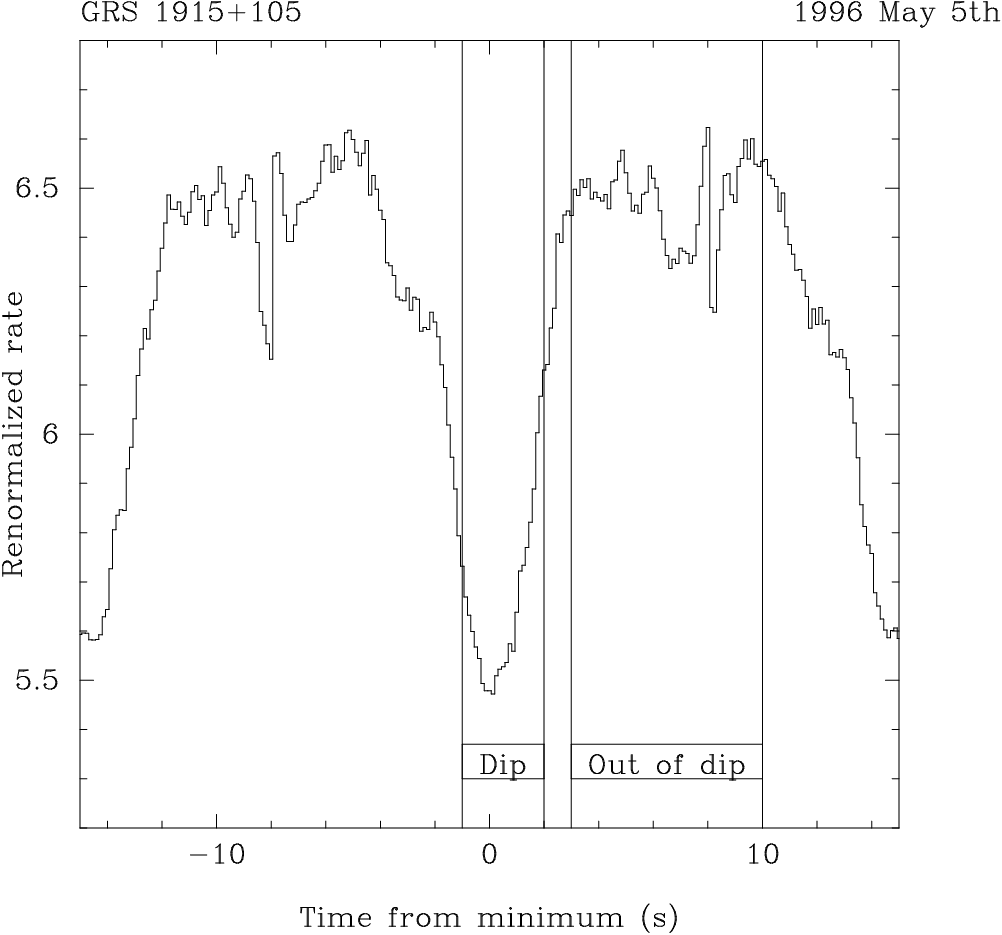}
  \caption{Folded light curve for the May 5th observation
          obtained by aligning the minima (see text). The marked bands are
          those used for the production of the `dip' and 'out of dip' PDSs.
	  }
  \label{fold_05}
\end{figure}

MRG97 report the detection of the 67Hz QPO
in 6 observations, all made in the first half of 1996. 
Some of the detections have a very low significance.
All these observations, with the exception of the first (Apr 6th),
correspond to class $\gamma$ in the classification by Belloni et al. (2000).
This class is indeed characterized by oscillations on a $\sim$10s time scale
(see Fig. 9 in MRG97),
oscillations that are accompanied by spectral changes (softening during the
narrow dips). In order to analyze in detail
whether the high-frequency QPO shows variations within one observation, 
a strong QPO detection is needed. Moreover, in order to study its possible
connection with the low-frequency oscillation ($\sim$100 mHz, see MRG97), 
it was necessary to obtain a good template for the latter. For this
purpose, the low-frequency oscillation needs to be
rather regular. Only one observation meets both requirements, that of May
5th, which was also analyzed by Cui (1999), who concentrated on the phase
lags of the different oscillations.
In addition, we analyzed another of those observations, May 14th, where the
67 Hz QPO over the whole observation has a low significance (2$\sigma$
in MRG97), but the low-frequency oscillation is rather regular.

\subsection{The May 5th observation}

\begin{figure}
\centering
\includegraphics[width=0.5\textwidth]{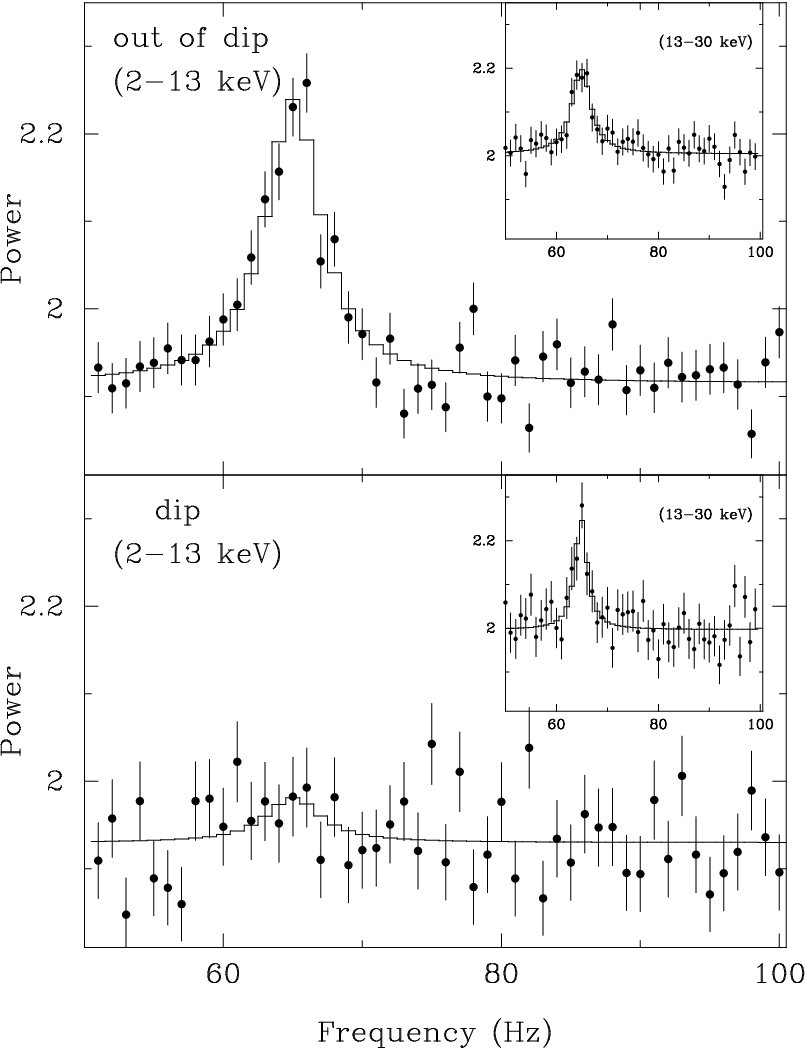}
\caption{High-frequency part of the PDSs obtained accumulating 
	'out of dip' (upper panel) and 'dip' (lower panel) intervals
        (2-13 keV). 
        The lines indicate the best fit models (see text).
	In the smaller panels, the corresponding PDSs for the 
	13-30 keV band are shown.
            }
  \label{pds_05}
\end{figure}
For each of the two observations, in order to identify the times of dips,
we followed a procedure similar to that 
adopted by MRG97. First we produced a light curve in the 2-13 keV band,
with a bin size of 0.125 s. Then we ran a sliding window with a 16s width
on the light curve. For each window, we fitted to the data a model consisting
of a constant plus a negative Gaussian profile centered on the middle of the 
window. For each fit, free parameters were therefore the constant level, and 
width and normalization of the Gaussian. As the next step, 
we identified the times
of the dips as local minima of the best fit amplitude of the Gaussian
amplitude in a 5-second time window, with the additional constraint that 
the amplitude had to be less than -1000 cts/s. Visual inspection of the light
curve showed that this was a reasonable way to identify the minima, which
are relatively easy to see by eye. A few spurious minima were deleted by
hand, the level of -1000 cts/s was chosen to be conservative.
With the minima, we constructed a template light curve by summing the 
data centered at the times of the minima.
The resulting template for May 5th can be seen in Fig. 1. The spikes at
-8s and +8s are spurious, corresponding to the limits of the window used
for the fits. The curve is similar to that in MRG97, although their procedure 
was slightly different: the oscillation is highly non-sinusoidal, being
constituted by a sharp dip every $\sim$15 seconds.
\begin{table*}
   \caption[]{Best fit parameters for the PDSs showed in Fig. 2 and Fig. 4.
        Upper limits are at
	95\% confidence. (FIX) indicates that a parameter has been 
	fixed to the quoted value (see text).}
       \label{tab1}
   $$
   \begin{array}{lcccccc}
     \hline
     \noalign{\smallskip}
     
{\rm Interval}  &\nu_0 (Hz)   & \Delta (Hz) & \%{\rm rms}         & \chi^2 (dof) & {\rm HR}_1 & {\rm HR}_2\\
     \noalign{\smallskip}
     \hline
     \noalign{\smallskip}
\multicolumn{7}{c}{{\rm 1996~May~5th (2-13 keV)}}  \\
     \noalign{\smallskip}
     \hline
     \noalign{\smallskip}
{\rm Out~of~dip}&65.01\pm 0.17 & 4.69\pm 0.45 & 4.32 \pm 0.20 & 51(45) & 1.2248 \pm 0.0003 & 0.1245 \pm 0.0001\\
{\rm Dip}       &65.01 {\rm (FIX)}    & 4.69 {\rm (FIX)}    &  < 2.49 (95\%)& 53(47) & 1.1773 \pm 0.0005 & 0.1248 \pm 0.0002\\
     \noalign{\smallskip}
     \hline
     \noalign{\smallskip}
\multicolumn{7}{c}{{\rm 1996~May~5th (13-30 keV)}}  \\
     \noalign{\smallskip}
     \hline
     \noalign{\smallskip}
{\rm Out~of~dip}&64.87\pm 0.27 & 4.41\pm 0.84 & 5.52 \pm 0.42 & 37(45)\\
{\rm Dip}       &64.81\pm 0.24 & 2.74\pm 0.87 & 5.56 \pm 0.56 & 37(45) \\
     \noalign{\smallskip}
     \hline
     \noalign{\smallskip}
\multicolumn{7}{c}{{\rm 1996~May~14th (2-13 keV)}} \\
     \noalign{\smallskip}
     \hline
     \noalign{\smallskip}
{\rm Dip}       &27.09 \pm 0.36 & 9.08 \pm 1.15 & 2.24 \pm 0.15 & 94(84) & 1.1432 \pm  0.0005 & 0.1459 \pm 0.0002\\
{\rm Out~of~dip}&27.09 {\rm (FIX)}    & 9.08 {\rm (FIX)}    &  < 0.85 (95\%)& 140(86)& 1.0626 \pm  0.0004 & 0.1470 \pm 0.0002\\
     \noalign{\smallskip}
     \hline
\end{array}
   $$
\end{table*}

From our template, we identified two intervals: a 3s interval around the 
minimum, corresponding to the `dip', and a 7s interval where the flux
is at its maximum and rather constant level (see Fig. 1).
The 3s dip interval is not symmetric around the minimum since from the
procedure for the extraction of PDSs (see below), we were limited to
intervals 1 second long. A selection with the dip interval shifted 
backwards by 1 second gave compatible results.
We then divided the same 2-13 keV light curve in intervals 1 second long
and accumulated a Power Density Spectrum (PDS) 
for each of the intervals, at the highest resolution
available (2 ms), corresponding to a Nyquist frequency of 256 Hz). 
The PDSs were normalized according to Leahy et al. (1983).
Finally, we summed all the PDSs corresponding to the `dip' and `out of dip' 
times described above to obtain the average PDS for each of the two intervals.
The results can be seen in Fig. 2. It is evident that the QPO at $\sim$65 Hz
is present only in the high-flux PDS. We fitted the `out of dip' PDS in Fig. 2 
with a model consisting of a Lorentzian (with centroid frequency $\nu_0$ and
FWHM $\Delta$) for the QPO peak and a constant
to take into account the contribution by the Poissonian noise. The best fit
value for the QPO power has been converted to fractional rms by taking into
account the net source rate, after subtraction of the average background rate
estimated with the FTOOL {\tt pcabackest}. The best
fit parameters can be seen in Table 1. The QPO peak at 65 Hz 
has a fractional rms amplitude of 4.32$\pm$0.20\% (significant at a
11.3 $\sigma$ level) in the 2-13 keV energy range.
For fitting the `dip' PDS, we fixed centroid and width of the
Lorentzian to the `out of dip' values and obtained a 95\% upper limit on the
fractional rms of the QPO of 2.49\%, significantly lower than the detection 
outside the dips.

\begin{figure}
\centering
\includegraphics[width=0.5\textwidth]{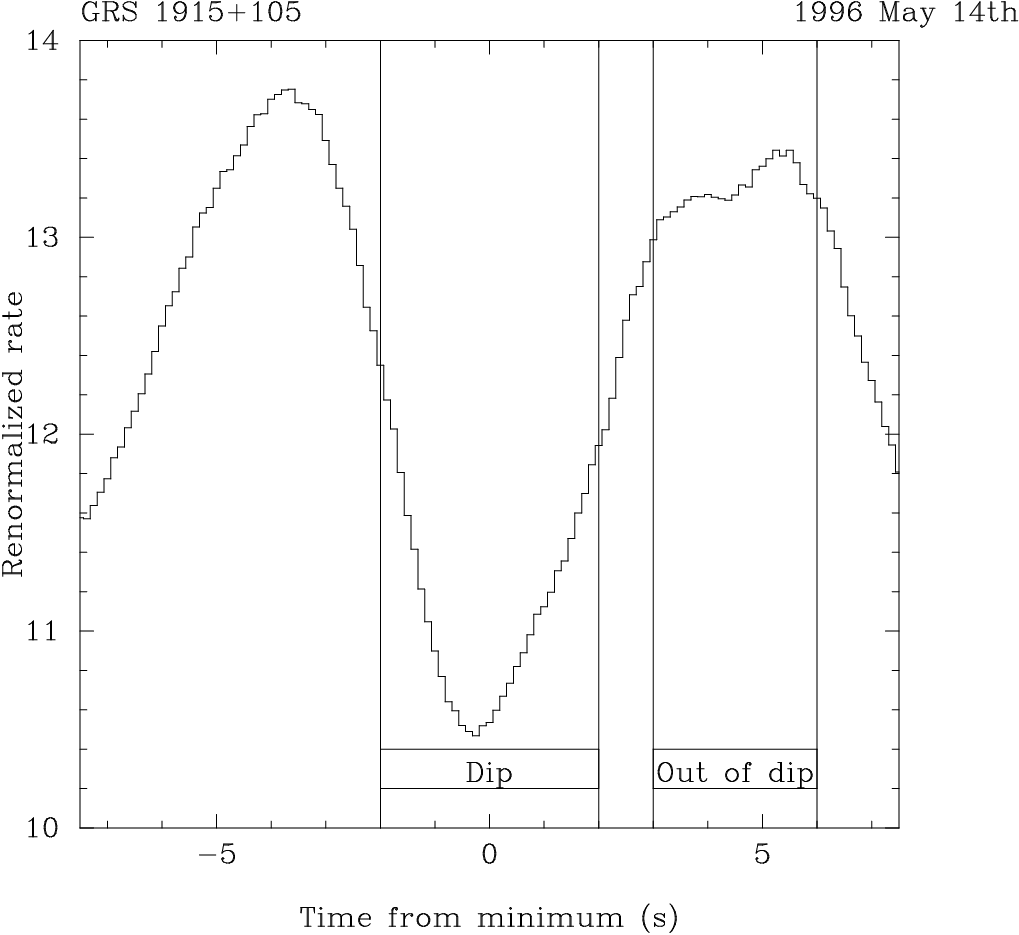}
\caption{Folded light curve for the May 14th observation
	obtained by aligning the minima (see text). The marked bands are
	those used for the production of the `dip' and 'out of dip' PDSs.
	    }
  \label{fold_14}
\end{figure}

Since in MRG97 the fractional rms of the 65 Hz QPO increases with increasing
energy, we produced power density spectra also for a higher energy band
(13-30 keV). In this band, the 65 Hz QPO is detected both in the dips and
outside (5.0$\sigma$ and 6.5$\sigma$ respectively, see small panels in 
Fig. 2), 
with a centroid frequency compatible with that of the softer energy band, with
a fractional rms of $\sim$5.5\%, similar to that reported by MRG97 (see
Table 1).
The only difference here is that the QPO peak is
narrower during the dips (2.74$\pm$0.87 Hz) than outside (4.41$\pm$0.84 Hz).

\subsection{The May 14th observation}

We repeated the same procedure for the May 14th observation. Our hope was
to be able to extract a more significant detection for the 65 Hz QPO by
de-selecting the dip intervals. As already shown by MRG97, 
the low-frequency oscillation in this observation is faster ($\sim$ 9s),
and its shape is roughly sinusoidal, therefore different from the sharp
dips shown before. This can be seen in the template shown in Fig. 3.
Nevertheless, we selected a 4s stretch around the minimum, which we call
for consistency `dip', and a 3s stretch around the maximum as 
`out of dip' (see Fig. 3).
The resulting averaged PDSs (2-13 keV) can be seen in Fig. 4. 
No significant peak
is present around 65 Hz in either spectrum, but a broad peak around 27 Hz
is visible in the `dip' spectrum. This peak is easily seen also in the total
PDS shown by MRG97.

Again, we fitted the PDSs, adding to the model described above a power law
to describe the low-frequency excess, which we could not simply
window out as in the previous case. The parameters are shown in Table 1.
In the `dip', the QPO peak at 27 Hz has a fractional rms amplitude of 
2.24$\pm$0.15\% (significant at the 7.5$\sigma$ level).
Outside the dips, fixing the QPO parameters to the `dip' ones, we obtained
a 95\% upper limit on the fractional rms of a 27 Hz peak of 0.85\%.

\begin{figure}
\centering
\includegraphics[width=0.5\textwidth]{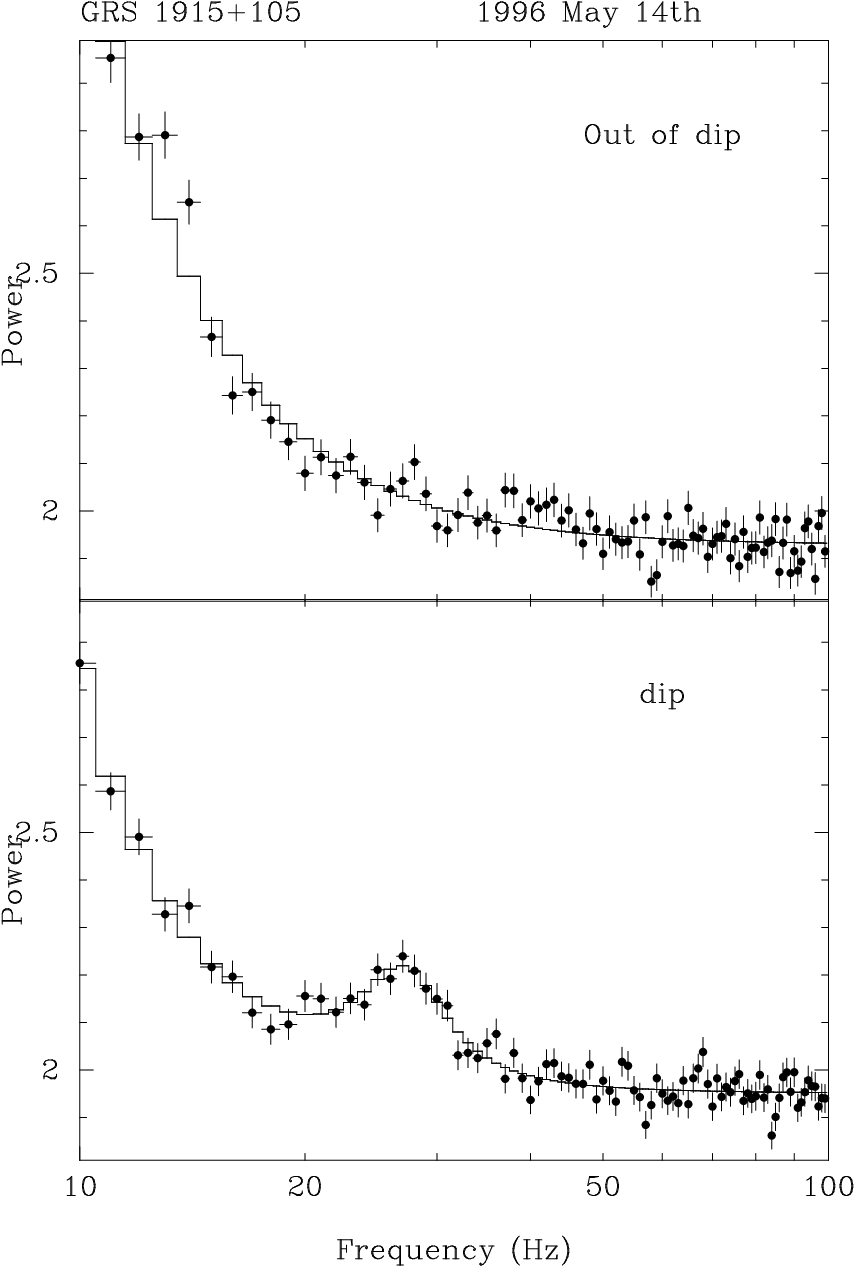}
\caption{High-frequency part of the PDSs obtained accumulating
        'out of dip' (upper panel) and 'dip' (lower panel) intervals
	(2-13 keV).
        The lines indicate the best fit models (see text).
            }
   \label{pds_14}
\end{figure}

As in the previous case, we repeated the analysis in the 13-30 keV band, 
but found no QPO at the same frequency. The derived upper limits are not
very stringent and completely compatible with the low-frequency detection
(2.9\% and 4.5\% for the out-of-dip and dip respectively).

\subsection{Spectral variations during the dips}

In order to check whether the observed dips are indeed transitions 
from state B to 
state A of Belloni et al. (2000), since these two observations were not
included in their sample, we accumulated X-ray colors for the dip and 
out of dip intervals. We chose the same X-ray bands and the same
definitions of colors as in Belloni et al. (2000). Colors are defined
as HR$_1$=B/A and HR$_2$=C/A, where A, B and C are the net source counts
in the channel bands 0-13 (2-5 keV),14-35 (5-13 keV), and 36-255 
(13-60 keV) respectively. For each observation, all the dip and out-of-dip
data were summed to compute the colors.
The result is the same for both observations:
the soft color HR$_1$ decreases significantly during the dips, while the
hard color does not change within the errors for May 5th, and decreases
very slightly on May 14th. This behaviour is consistent with A--B transitions
(see Belloni et al. 2000).

To investigate more thoroughly the spectral variations of the May 5th 
observations, we extracted energy spectra from the PCA data, using the
high time resolution data modes ({\tt binned} and {\tt event} modes).
The spectral resolution of these modes for this observation is rather low, 
featuring only four independent channels below 13.5 keV, but it is important
to have an idea of the energy spectrum and to calculate the relative 
contribution of the disk and power-law component. We extracted the spectra
and the PCA response matrices using {\tt FTOOLS V5.0.1}. Before extracting
the spectra for our aims, we tested our extraction procedure by accumulating
a 96s spectrum both from high time resolution data and from standard data,
which have full energy resolution: we obtained compatible results.
We then accumulated
two spectra, corresponding to the dip and out-of-dip intervals used for
the production of the PDSs (see Sect. 2.2). We fitted both spectra
(excluding the first channel and the channels above 25 keV because 
of calibration uncertainties at low and high
energies, effectively restricting ourselves to the 5-25 keV interval),
with a model consisting of a disk-blackbody plus a power law (with an
artificial low-energy break at 4 keV in order to avoid a power-law excess
at low energies). The interstellar absorption was fixed to 6$\times 10^{22}$
cm$^{-2}$ (see Belloni et al. 2000). A 1\% systematic error was added
to all channels to account for residual calibration uncertainties.
The fits proved rather problematic: no good fit in the statistical sense 
could be obtained, with reduced $\chi^2$ values of 2.8 and 4.0 (6 dof)
for the out-of-dip and dip spectrum respectively.
At this stage, we are not able to provide a robust estimate of the parameter
changes that give rise to the difference in the observed colors. Overall, 
as expected (see Belloni et al. 2000), the disk has a rather high
temperature in both spectra ($\sim 2$ keV) and a small inner radius
($\sim 20$km), while the power law is steep ($\Gamma\sim$ 3.5).
However, given the relatively small changes of the spectral parameters
between the two spectra (the only significant change is the inner radius
of the disk component, larger in the out-of-dip spectrum, we can 
estimate roughly the percentage of detected photons
coming from the two components. 
In the out-of-dip spectrum,
the power law contributes 49\% of
the total observed counts in the low-energy band, and about the same 
percentage in the high-energy band. In the dip spectrum, these percentages
are 52\% and 50\% respectively. From the uncertainties in the spectral
parameters, we can estimate an error on these percentages of $\pm$3\%.

\section{Discussion}

The analysis presented in the previous section shows that there is a 
relation between the low-frequency oscillation and the presence of the
high-frequency QPO, on a time scale a factor of $\sim$1000 shorter.
However, this relation is complex and difficult to interpret. The energy
spectrum of black hole candidates in general and of GRS~1915+105 in 
particular can be decomposed into two separate components: a thermal
component thought to originate from an optically thick accretion disk
and a high-energy component associated with an optically thin flow (see
Tanaka \& Lewin 1995; Belloni et al. 1997a,1997b). 
The energy dependence of the fractional rms of the 65Hz QPO analyzed in 
this work is rather hard, indicating that it must be associated with the
optically thin component (MRG97). In the following, we will make this
assumption.

Let's examine the two different PDSs of the May 5th observation.

\begin{itemize}

\item {\it Out-of-dip}:  in this case, the fractional rms of the detected
	65 Hz QPO is 4.3\% in the soft energy band and 5.5\% in the 
	hard energy band.
        (see Sect. 2.2). These figures have been calculated
	by assuming the QPO is associated with the {\it total} flux
	from the source. If we make the assumption that the QPO comes
	only from the power-law component, using the fraction of
	counts from the power-law component from Sect. 2.4, the
	fractional rms in the soft energy band increases to 8.8\%,
	that in the hard energy band to 11.2\%.

\item {\it Dip}: here the QPO has an upper limit of 2.5\% in the
	soft band, and a detection of 5.5\% in the hard band.
	Using the figures from Sect. 2.4, under the same assumption
	that the QPO is associated only with the power-law
	spectral component, we can convert these
	numbers to 4.8\% and 11\% respectively.

\end{itemize}

From this exercise, we can conclude a few things. The first is that 
the increase in the fractional rms of the QPO reported by MRG97,
associated with the out-of-dip intervals, could be entirely due to the
fact that the QPO is associated with the hard power-law component and
at low energies is diluted by the disk component. The second is that
we can exclude that the non-detection of the QPO in the soft
band during the dips
is due to a relative increase of the contribution of a non-variable
disk component. This because after correcting for the power-law fraction
(see above), the soft-band detection in the out-of-dip data is
8.8$\pm$0.7\% and the 95\% upper limit in the dip data is 4.8\%.
An alternative way to estimate this effect is to calculate by how
much the power-law fraction would have to decrease in the dip spectrum
with respect to the out-of-dip spectrum in order to make the detection
in the one compatible with the upper limit in the other. The resulting
number is 0.58, while within the errors the two fractions remain the
same.
Notice that also the amplitude of the low-frequency
oscillation increases with energy (Tomsick \& Kaaret 2001), as do most
of the QPOs observed in black hole candidates (see van der Klis 1995).

Unfortunately, a more detailed analysis is not possible due to the lack
of energy resolution in the high time resolution energy spectra. However,
our results show that the presence of the 65 Hz QPO, which has been 
reported only for a limited number of observations, is indeed limited to
a very specific state of the source, which we identify with state B
of Belloni et al. (2000). The periodic fast excursions to state A (the
dips) are already enough  to produce sharp changes in the properties
of the QPO, whose spectrum must change in order not to have it 
detected in the soft energy band. A thorough analysis of a large number
of PCA observations of GRS~1915+105 is needed, but is clearly beyond the
scope of this work.

Finally, the PDSs of May 14th show that we are far from having a complete
picture of the timing properties of GRS~1915+105: here the 65 Hz does not
seem to be significantly present in either the out-of-dip or the dip
PDSs, but another QPO at 27 Hz appears, only during the dips. Not much
more can be said about this, in the lack of additional datasets showing 
oscillations at these and possibly other frequencies. Of course, given
the peculiar property of GRS~1915+105 to change its visible inner disk
radius over short time scales, none of the models proposed for the 
generation of a high-frequency QPO (see e.g. MRG97; Nowak et al. 1997; 
Cui et al. 1998) are ruled out by these results.

\begin{acknowledgements}
We thank S. Campana for useful discussions.
TB thanks the Cariplo foundation for financial support.
This work was supported by the Netherlands Research School for Astronomy
(NOVA), the Netherlands Organization for Scientific Research (NWO) under
contract number 614-51-002 and the NWO Spinoza grant 08-0 to E.P.J. 
van den Heuvel. MM is a fellow of the Consejo Nacional de Investigaciones
Cient\'{\i}ficas y T\'ecnicas de la Rep\'ublica Argentina. 
CSF was partially supported by a `Rafael Calvo Rodes' fellowship
from Instituto Nacional de Tecnica Aeroespacial, Spain.
This research
has made use of the data obtained through the High Energy Astrophysics
Science Data Archive Research Center Online Service, provided by the 
NASA/Goddard Space Flight Center.
\end{acknowledgements}

\end{document}